\documentclass[10pt,letterpaper]{article}
\usepackage[utf8]{inputenc}
\usepackage{amsmath}
\usepackage{amsfonts}
\usepackage{amssymb}
\usepackage{graphicx}
\usepackage[left=2cm,right=2cm,top=2cm,bottom=2cm]{geometry}
\usepackage{tikz, xcolor, colortbl, hyperref}
\usepackage{url}
\usepackage{float}
\usepackage[procnumbered, linesnumbered, ruled, vlined, lined]{algorithm2e}
\usepackage{setspace} 
\usepackage[affil-it]{authblk}

\newtheorem{lemma}{Lemma}

\definecolor{lime}{HTML}{A6CE39}
\DeclareRobustCommand{\orcidicon}{%
	\begin{tikzpicture}
	\draw[lime, fill=lime] (0,0) 
	circle [radius=0.16] 
	node[white] {{\fontfamily{qag}\selectfont \tiny ID}};
	\draw[white, fill=white] (-0.0625,0.095) 
	circle [radius=0.007];
	\end{tikzpicture}
	\hspace{-2mm}
}

\foreach \x in {A, ..., Z}{%
	\expandafter\xdef\csname orcid\x\endcsname{\noexpand\href{https://orcid.org/\csname orcidauthor\x\endcsname}{\noexpand\orcidicon}}
}

\title{An Algorithm for the Euclidean Bounded Multiple Traveling Salesman Problem}
\author[1]{Víctor Pacheco-Valencia \orcidA{}}
\author[1]{Nodari Vakhania \thanks{Corresponding author. E-mail: nodari@uaem.mx} \orcidB{}}

\affil[1]{Centro de Investigación en Ciencias\\
          Universidad Autónoma del Estado de Morelos\\
          Cuernavaca, Morelos, México}

\begin{document}

\maketitle

\abstract{
In the Bounded Multiple Traveling Salesman Problem (BMTSP), a tour for
each salesman, that  starts and ends at the depot and that respects the
bounds on the number of cities that a feasible salesman tour
should satisfy, is to be constructed. The objective is to minimize the
total length of all tours. Already Euclidean traveling salesman
problem is $NP$-hard. We propose a 3-Phase heuristic algorithm for the
Euclidean BMTSP. We tested the algorithm for the 22 benchmark instances and
168 new problem instances that we created. We report 19 best known 
solutions for the 22 benchmark instances including the 12 largest ones.
For the newly created instances, we compared the performance of our 
algorithm with that of an ILP-solver CPLEX, which was able to construct 
a feasible solution for 71\% of the instances within the time limit of 
two hours imposed by us. 
For about 10\% of the smallest new instances, CPLEX delivered slightly 
better solutions, where our algorithm took less than 180 seconds for the largest 
of these instances.  For the remaining 61\% of the instances solved by CPLEX, 
the solutions by our heuristic were, on average, about 21.5\% better 
than those obtained by CPLEX.
}

\setlength{\parskip}{15pt}

\section{Introduction}
The Multiple Traveling Salesman Problem (MTSP) is an extension of a well-known
Traveling Salesman Problem (TSP) in which these are $k\ge 1$ salesmen, instead
of one salesman in TSP. The Bounded Multiple Traveling Salesman Problem (BMTSP),
in turn, is an extension of MTSP with an additional 
restriction on the number of cities that each salesman has to visit. This
kind of restriction is important in most real-life applications 
since each salesman has a fixed working hours within which a restricted 
number of cities can be visited, where the employer wishes to balance
the load of each salesman. Hence, the lower and upper bounds $m_{\min}$
and $m_{\max}$ on the number of cities that each salesman should visit, become 
problem parameters. We have $n+1$ cities, among which there is one 
distinguished city $d$ called the {\em depot}, and there are $k$ salesmen. A feasible
tour of a salesman starts in depot, visits at least  $m_{\min}$ and at most 
$m_{\max}$ cities, each one exactly once, and returns to the depot. We are
given the distance (the transportation cost) between a pair of cities. Here, we
consider the Euclidean setting where the distances are naturally  
determined in a 2-dimensional Euclidean space. The cost of a tour is the
sum of the distances between the cities in that tour. The objective is to 
form $k$ tours satisfying the just stated restrictions to minimize 
the total cost of all $k$ tours. 

BMTSP  takes into account natural limits on the number of visited cities for 
each salesman by imposing an upper bound $m_{\max}$. At the same time, 
a lower bound $m_{\min}$   ``controls'' the minimal load for each 
salesman  \cite{roerty1974,okonjo1988,garn2021}. In practical terms, 
an optimal solution provides highest profit for an involved company,  
customer satisfaction and implies less contamination of the environment.

Already Euclidean TSP is known to be $NP$-hard \cite{papadimitriou1977}. 
Among the particular applications, we mention those in 
(1) transportation and delivery processes, where ground and air vehicles combine to optimize 
package deliveries  \cite{angel1972,svestka1973,lenstra1975,zhang1999,zhang2018,ibrocska2023}\sloppy;
(2) wireless sensor networks  where mobile robots and drones collect  
data from ground sensors for network connectivity  \cite{saleh2004,zhan2019,venkatesh2015}; 
(3) search, rescue and disaster management, cooperative mission, 
monitoring and surveillance operations,  where routes and 
rescue efforts are to be optimized  \cite{brumitt1996,calvo2003,faigl2005,ergezer2014,yanmaz2023,manshadian2023}; 
(4) precision agriculture, where mobile 
robots and drones are used to improve efficiency and productivity \cite{du2017, dolias2022, manshadian2023};
(5) general multi-robot task allocation and scheduling problems, where task 
assignment to robots is to be optimized \cite{gorenstein1970,gilbert1992,tang2000,yu2002,sariel2009,carter2009,pacheco2022-overview}. In 
pragmatic terms, quality solutions to BMTSP offer benefits to workers, customers, 
society in general, and the whole community as they help to 
reduce operating costs \cite{garg2023} and risks  \cite{manshadian2023},
improve process efficiency, worker and customer satisfaction and timely package 
deliveries \cite{garg2023}. All of this encourages customer loyalty and 
increases sales while increasing also work efficiency. By reducing fuel consumption, 
we effectively reduce greenhouse gas emissions, air pollution and worker stress \cite{rondinelli2000,tanczos2008,garg2023}. 

Next, we give a brief look  at the existing literature on the solution 
methods for MTSP and BMTSP. We refer the reader to \cite{bektas2006}
for an extensive overview of the existing approaches for MTSP and its
variations, including the one with more than one depot, the bounded MTSP, 
MTSP  with time windows and with a variable number of salesmen. For a more 
recent survey with a focus on some real-life applications, see \cite{cheikhrouhou2021}. \cite{pierce1969}, 
\cite{gorenstein1970}, 
\cite{bellmore1974}, 
\cite{lenstra1975}, 
\cite{russell1977}, 
\cite{hong1977}, 
\cite{rao1980}, 
\cite{jonker1988}, and 
\cite{potvin1989} 
use a transformation of MTSP to  TSP to obtain approximation solutions. 
As to the exact methods, \cite{miller1960} 
proposed an integer linear programming (ILP) formulation for the BMTSP imposing 
only an upper bound $m_{\max}$ on the number of cities in each tour. 
\cite{svestka1973} 
proposed an ILP formulation for MTSP and a branch-and-bound
algorithm, that was able to solve instances with up to 60 cities. Later
\cite{gavish1976}  has corrected the ILP from  \cite{svestka1973}.
\cite{laporte1980} 
proposed ILP formulations for symmetric and also asymmetric MTSP  
with a variable number of salesmen and a common fixed cost $f$ for each salesman.
The proposed straight and inverse algorithms were able to solve 
optimally instances with up to 100 cities and with up to 10 salesmen. 
\cite{christofides1981} 
proposed two ILP formulations for the capacitated MTSP, the first one for 
the uncapacitated symmetric MTSP, which was used to derive a lower bound for the
capacitated version. 
\cite{ali1986} 
proposed a branch-and-bound algorithm for  MTSP that obtained
optimal solutions for asymmetric instances with sizes up to 100 cities
and symmetric Euclidean version with up to 59 cities. 
\cite{gavish1986} 
made an attempt to solve larger scale symmetric non-Euclidean MTSP instances with up 
to 500 cities with $k=2,4,5,6,10$ salesmen, and Euclidean instances with up to 100 
cities and 10 salesmen with branch-and-bound.
\cite{kara2006} 
presented an ILP formulation for BMTSP with both, lower and upper bounds 
$m_{\min}$ and $m_{\max}$, and for a few extensions of BMTSP. Using an ILP-solver
CPLEX, they were able to solve 
18 instances in an average processing time of 13.06 seconds for $n=30$, $k=3,4,5$, $m_{\min} \in [3,5]$ and $m_{\max} \in [6,15]$; 
18 instances  in an average processing time of 158.33 seconds for $n=50$, $k=3,4,5$, $m_{\min} \in [3,5]$ and $m_{\max} \in [10,25]$; and 
18 instances in an average processing time of 747.17 seconds for $n=70$, $k=3,4,5$, $m_{\min} \in [3,5]$ and $m_{\max} \in [15,30]$. 

Heuristics and meta-heuristics have been used to solve large-sized MTSP and BMTSP instances. A genetic algorithm  proposed in  
\cite{tang2000}, 
modeled BMTSP for hot rolling scheduling problem in a Steel Complex.  
\cite{carter2002} 
  modeled the production process of pre-printed newspaper advertisements using 
  solutions to MTSP. 
Later \cite{carter2006} 
 modified their algorithm and used it to solve 
 12 Euclidean MTSP instances with $n=51$ for $k=3,5,10$; $n=100$ for $k=3,5,10,20$; and $n=150$ for $k=3,5,10, 20, 30$. 
\cite{brown2007}, 
\cite{singh2009} and 
\cite{yuan2013}, 
 gave alternative genetic methods considering two objective functions, the maximum 
 length of a salesman tour and the total tour length, and applied them to the 12 instances from  \cite{carter2006}. 
\cite{junjie2006}, 
 created 6 Euclidean two-dimensional BMTSP instances with $n=76,152,226,299,439,1002$ 
 for $k=5$, $m_{\min}=1$ and $m_{\max}=20,40,50,70,100,200$ respectively.
 They compared the solutions obtained by their ant colony optimization algorithm 
 with those from \cite{tang2000}. For the three largest instances, they obtained
 better solutions, while the algorithm from \cite{tang2000} was better for the three remaining smallest instances. 
\cite{sedighpour2012}, 
solved the 6 benchmarks instances from \cite{junjie2006} using an alternative genetic algorithm improving earlier best known solutions (BKS) for four of them.
\cite{yousefikhoshbakht2013} 
tested their alternative ant colony optimization algorithm for the 6 benchmarks 
instances from  \cite{junjie2006} improving the  BKS for 5 of them. 
\cite {larki2014}, 
proposed a hybrid algorithm for MTSP combining an imperialist competitive algorithm with the Lin-Kernigan algorithm \cite{lin1973}. \cite{venkatesh2015}
 proposed an alternative genetic method for MTSP considering two  objective functions, 
 the maximum length of a salesman tour and the total tour length. 
\cite{necula2015} 
 created 16 new Euclidean instances for BMTSP with 
 $n=51,52,76,99$ and $k=2,3,5,7$ and suggested a few ant colony algorithms, 
 which they compared with the results obtained by solving the corresponding ILP model 
 by CPLEX. Curiously, CPLEX delivered better
 solutions for all these instances, among which there were 6 optimal ones. 
\cite{lo2018} applied their genetic algorithm 
 to  the 6 BMTSP instances from  \cite{junjie2006} improving earlier BKS for all
 these instances. 
A hybrid genetic algorithm proposed by \cite{alfurhud2020} 
 improved 5 BKS for the 6 BMTSP benchmarks instances by \cite{junjie2006}
A three phase heuristic described in \cite{pacheco2021-ioca} 
was tested for the existing 22 Euclidean BMTSP benchmark instances, and improved the BKS
for the largest of them. 



The above mentioned 22 benchmark instances were created using a simple selection of 
the parameters  $k$, $m_{\min}$ and $m_{\max}$. For instance, the five largest ones 
from \cite{junjie2006} have a fixed value of $k=5$ and $m_{\min}=1$ 
with $m_{\max} \approx  n/{4.5}$. 
Inspired by practical scenarios, we created 168 new instances. We 
derived the new instances from the earlier known TSP instances from 
TSPLIB \cite{TSPLIB-Benchmarks}, where the parameters $k$, $m_{\min}$ and $m_{\max}$ 
were created based on the statistics obtained from a number of real-life 
applications reflecting the sizes of delivery vehicle fleets currently 
in use, they also have realistic number of salesmen and realistic ranges 
$[m_{\min},m_{\max}]$ (see Section \ref{sResults} for the details). In our
instances, the parameter $k$ ranges from 2 to 69, where it ranges from 2 to
7 for the earlier benchmark instances. The newly created 
instances contain 30 larger sized instances with up to 1577 cities 
with larger values for both $n$ and $k$ compared to the earlier benchmark 
instances. The newly created problem instances, 
together with the code of our algorithm and the detailed description of our results
are publicly available  at  \cite{pacheco2023-benchmarks}. 

The heuristic, the Partition, Construction and Improvement (PCI) algorithm,
that we propose here,  has the worst-case time complexity $O(k^2 n^4)$ and  
runs in three phases.    
Phase 1 partitions the set of cities in $k$  subsets, phase 2  builds a feasible 
tour for each of these subsets, and phase 3 improves iteratively the $k$ tours
using hill climbing techniques. 
We use CII algorithm  \cite{pacheco2020-cii} to solve each of the $k$ TSP 
instances to obtain $k$ feasible tours. (The CII-algorithm with the worst-case time 
complexity of $O(n^2)$, designed for the Euclidean TSP, proved to give good 
approximation and to be very fast in practice \cite{pacheco2020-cii}). 

A brief description of our experimental study is as follows. For 
the 22 benchmark BMTSP instances, PCI-algorithm improved the earlier BKS 
for 15 of them, including the 12  largest ones,
and it obtained the earlier BKS for 4 instances. Since no earlier statistics 
for the newly generated 168 instances were available, we used an ILP-solver 
CPLEX   \cite{cplex2024} to judge about the quality of 
our solutions for these instances.  CPLEX was able to obtain 119 solutions, 
one of them being optimal, within the time limit of 2 hours that was imposed by us. 
Among these solutions, 17 turned out to be, on average, about 0.8\% better  than 
the corresponding solutions from our heuristic.  For the remaining 102 instances, 
the solutions obtained by PCI-algorithm were, on average, about 21.5\% better than 
ones from CPLEX. As to the execution times, for the 119 instances solved by CPLEX, 
our algorithm was, on average, about 613 times faster.  
It is remarkable that the solution obtained by PCI-algorithm 
for the smallest of the former 12 instances, was, in turn, improved by 
CPLEX. PCI-algorithm was faster than the earlier reported 
best execution times for all benchmark instances except one (the best time
for the latter instance was obtained by the algorithm from  \cite{pacheco2021-ioca}).
PCI-algorithm delivered solutions in less than one minute for about 81\% of 
the 22 benchmark instances. 

The rest of this text is organized as follows. In the next section we give
a formal description of our problem and its Integer Linear Programming formulation.
The next three  sections describe the three phases of our algorithm.
Section \ref{sPartition} describes the partitioning procedures used in Phase 1.
Subsection \ref{sConstruction} describes the procedure of Phase 2 that constructs 
a feasible tour. Section \ref{sImprovement} contains our improvement procedures. 
Some results from this work, in a preliminary form,  were presented at
\cite{pacheco2023-india}.

\section{Problem definition}
\label{sMILP}

First, we define BMTSP.
We let  $w(i,j)$  be the cost of a salesman's travel from city 
$i$ to city $j$.  Each city from set $V\setminus \{d\}$ is to be visited 
by exactly one salesman. A feasible tour $T^l$ of salesmen $k$, 
\begingroup
\small
\setlength{\abovedisplayskip}{1pt}
\setlength{\belowdisplayskip}{1pt}
\begin{equation}
T^l = (d, i_1^l, i_2^l, \cdots, i_{m_l-1}^l, i_{m_l}^l, d)
\end{equation}
\endgroup
starts from the depot, 
ends in the depot and visits at least  $m_{\min}$ and at most $m_{\max}$ cities, 
exactly once each of them; here $\{i_1^l,\dots i_{m_l}^l\}$ is the set of cities 
that salesman $l$  visits in tour $T^l$,  $m_{\min}\le m_l\le m_{\max}$.

The total cost or the cost of tour $T^l$, denoted by $C(T^l)$, is the sum of 
the costs between each pair of cities along that tour:
\begingroup
\small
\setlength{\abovedisplayskip}{1pt}
\setlength{\belowdisplayskip}{1pt}
\begin{equation}
C(T^l)= w(d,i_1^l) + w(i_1^l, i_2^l) + \cdots + w(i_{m_l-1}^l, i_{m_l}^l) + w(i_{m_l}^l,d)
\label{eCTj}
\end{equation}
\endgroup
A feasible solution (tour) $T=\{ T^1, T^2, \cdots , T^k\}$ 
includes the $k$ salesmen tours and has the cost 
\begingroup
\small
\setlength{\abovedisplayskip}{1pt}
\setlength{\belowdisplayskip}{1pt}
\begin{equation}
C(T)=C(T^1)+C(T^2)+\cdots+C(T^k)
\label{eCT}
\end{equation}
\endgroup

The objective is to find an optimal solution (tour), a feasible one with the 
minimum cost $\min_{T} \ C(T)$.

Next, we give an ILP formulation of BMTSP, 
see \cite{kara2006} for the details. 
Let $G(V,E)$ be a complete, directed weighted graph that defines an 
instance of BMTSP. The set of vertices, $V$, consists of $n$ elements with
indices $2, \dots, n+1$ and a the depot identified with index $1$. We have $(n+1)^2$ 
edges, denoted by $(i,j) \in E$, ($i,j \in V$),  with weights $w_{ij}=w(i,j)$ 
(representing the corresponding distances). In a given solution, for each $i,j \in V$,  
the binary variable  $x_{ij}$ is $1$ if the edge $(i,j)$ is  in the solution 
and it is $0$ otherwise. For a given solution, for each $i \ge 2$, an integer variable 
$u_i$ is the number of vertices before vertex $i$ in the tour of that solution to which 
vertex $i$ belongs. 

The ILP formulation is as follows: 
\begingroup
\small
\setlength{\abovedisplayskip}{1pt}
\setlength{\belowdisplayskip}{1pt}
\begin{equation}
\label{eKara-one}
\text{minimize} \quad \sum_{i=1}^{n+1} \sum_{j=1}^{n+1} w_{ij} x_{ij}
\end{equation}
\endgroup
Subject to the following restrictions:
\begingroup
\small
\setlength{\abovedisplayskip}{1pt}
\setlength{\belowdisplayskip}{1pt}
\begin{alignat}{1} 
\label{eKara-two}
&\sum_{j=2}^{n+1} x_{1j} = k\\
\label{eKara-three}
&\sum_{i=2}^{n+1} x_{i1} = k\\
\label{eKara-four}
&\sum_{i=1}^{n+1} x_{ij} = 1, \qquad \text{for} \ j=2, \dots, n+1\\
\label{eKara-five}
&\sum_{j=1}^{n+1} x_{ij} = 1, \qquad \text{for} \ i=2, \dots, n+1
\end{alignat}
\endgroup

\begingroup
\small
\setlength{\abovedisplayskip}{1pt}
\setlength{\belowdisplayskip}{1pt}
\begin{alignat}{1} 
\label{eKara-six}
&u_i + (m_{\max}-2)x_{1i}-x_{i1} \le m_{\max}-1, \qquad \text{for} \ i=2, \dots, n+1\\
\label{eKara-seven}
&u_i + x_{1i} + (2-m_{\min})x_{i1} \ge 2, \qquad \text{for} \ i=2, \dots, n+1\\
\label{eKara-eight}
&x_{1i} + x_{i1} \le 1, \qquad \text{for} \ i=2, \dots, n+1\\
\label{eKara-nine}
&u_i - u_j + m_{\max} x_{ij} + (m_{\max}-2)x_{ji} \le m_{\max} -1, \quad \text{where} \ 2 \le i \ne j \le n+1\\
\label{eKara-ten}
&x_{ij} \in \{0,1\}, \qquad \forall (i,j) \in E
\end{alignat}
\endgroup

Equation \eqref{eKara-one} is to minimize the solution cost. 
Equation \eqref{eKara-two} ( \eqref{eKara-three}, respectively) ensures that exactly $k$ 
tours start (end, respectively) in the depot. Equation \eqref{eKara-four} ensures that 
each city $j$ (except the depot) is entered exactly once, and equation \eqref{eKara-five}
ensures that there is exactly one tour that leaves each city $i$ (except the depot). 
Inequalities \eqref{eKara-six}and \eqref{eKara-seven} establish upper and lower limits
on the number of cities in each tour. Inequalities \eqref{eKara-eight} ensure that each
tour contains at least two cities. Inequalitiy \eqref{eKara-nine} ensures that there
are no two different tours containing vertices $i$ and $j$ (since $u_j=u_i+1$ if and 
only if $x_{ij}=1$ holds). 



\section{Phase 1: The partition}
\label{sPartition}

We now describe Phase 1 specifying how the partition of set $V\setminus \{d\}$ in 
$k$ subsets $V_1, V_2, \cdots, V_k $ is performed. The number of nodes in each of these subsets is in range $[m_{\min},m_{\max}]$). Phase 1 consists of 
stages 1 and 2. 

In Stage 1, an initial vertex is chosen for each of the subsets
$V_j$, $j=1, \cdots, k$ randomly. Then iteratively, every $V_j$ is extended with
a vertex, closest to one of the vertices already in $V_j$, $j=1, \cdots, k$, until
every set $V_j$ contains exactly $m_{\min}$ vertices (for none of these subsets
the depot is considered). 

The so formed partition is final if all vertices from set $V$ except the depot
are included in one of the subsets. Otherwise, stage 2 augments (some of) these
subsets as follows. Iteratively for every vertex 
$i \in V \setminus \{V_1 \cup \cdots \cup V_k \cup \{d\}\}$, the costs to
each vertex $x\in V_j$ with $|V_j| < m_{\max}$, $j=1, \cdots, k$, is calculated; 
vertex $i$ is included into any subset containing a vertex closest to $i$.  Phase
1 returns the resultant complete partition of the set $V$, $V_1,\cdots, V_k$.\\

\begin{spacing}{1.0}
\begin{procedure}[ht] 
\begin{scriptsize}
\DontPrintSemicolon

\For(\tcp*[f]{Stage 1}) {$j:=1$ \KwTo $k$}{
 $V_j:= \{ $ choose a vertex $i$ randomly from $V \setminus \{d\}$ with no repeats $\}$
}

\For{$m:=2$ \KwTo $m_{\min}$}{
 \For {$j:=1$ \KwTo $k$}{
  $V_j:=V_j \cup \{i \ | \ \min_{i \in V \setminus \{ V_1 \cup \cdots \cup V_k \cup \{d\}\}, \ l \in V_j} \{ w(i,l) \} \ \}$ \;
 }
}  

\While(\tcp*[f]{Stage 2}){$| V \setminus \{ V_1 \cup \cdots \cup V_k \cup\{d\}\}| >0 $} {
 $w^{\prime}:=\infty$\;
 \For {$j:=1$ \KwTo $k$}{
  \If {$|V_j| < m_{\max}$}{ 
   $i := \iota \ | \ \min_{\iota \in V \setminus \{ V_1 \cup \cdots \cup V_k \cup \{d\}\}, \ l \in V_j} \{ w(\iota,l) \}$\;
   $w:= w(\iota,l) \ | \ \min_{\iota \in V \setminus \{ V_1 \cup \cdots \cup V_k \cup \{d\}\}, l \in V_j} \{ w(\iota,l) \}$\;
   \If {$w < w^{\prime}$}{
    $i^{\prime}:=i$\;
    $w^{\prime}:=w$\;
    $j^{\prime}:=j$\;
   }
  }
 }
 $V_{j^{\prime}}:=V_{j^{\prime}} \cup \{ i^{\prime} \}$ \;
}  
\Return $V_1, \cdots, V_k$\;

\caption{$PARTITION$( ~)}

\label{pF1Partition}
\end{scriptsize}
\end{procedure}
\end{spacing}
\vspace*{1pt}


\begin{lemma}
The time complexity of Procedure \ref{pF1Partition} is $O(kn^3)$.
\label{lPartition}
\end{lemma}
\paragraph*{Proof}
In lines 1--2, the loop is repeated $k$ times, hence, the cost for 
this loop  is $O(k)$. 
In lines 3--5, the loop is repeated $m_{\min}-1$ times, and the loop in lines 4--5,  is repeated $(m_{\min}-1)k$ times.
In line 5, $(n-k)(m_{\min}-1)=(m_{\min}n-n-m_{\min}k+k)$ comparisons are needed 
to find the vertex $i$.
Hence, the time complexity for the loop of lines 3--5 is $O(m_{\min}^2kn)$.
In lines 6--16, the loop is repeated $n-m_{\min}k$ times.
In lines 8--15, the loop is repeated $(n-m_{\min}k)(k)=(kn-m_{\min}k^2)$ times.
In lines 10 and 11, the loop requires $(n-m_{\min}k)(m_{\max}-1)$ comparisons are 
needed to find the vertex $i$. 
Given that  $m_{\min}$  and $m_{\max}$ are bounded from the above 
by $n$, the computational cost for this loop can be expressed as $O(n^2)$.
Hence, the cost for the loop of lines 6--16 is $O(kn^3)$, 
and the time complexity of Algorithm \ref{pF1Partition} is 
$O(kn) + O(m^2_{\min}kn) + O(kn^3)=O(kn^3)$
(the second term transforms to $O(kn^3)$.

\section{Phase 2: Construction}
\label{sConstruction}

At Phase 2, we use a fast $O(n^2)$ algorithm for the TSP from \cite{pacheco2020-cii}. 
Thus a feasible tour $T^j$ for each of the subsets $V_j \cup \{d\}$ is constructed by 
merely invoking this algorithm, and  an initial feasible solution 
$T_0=\{ T^1, \cdots, T^k \}$ is returned. 

\begin{spacing}{1.0}
\begin{procedure}[ht]  
\begin{scriptsize}
\DontPrintSemicolon

$T_0:=\emptyset$\;
\For{$j:=1$ \KwTo $k$}{
 $T^j:=CII(V_j \cup \{d\})$\;
 $T_0:=T_0 \cup \{ T^j \}$\;
}

\Return $T_0$\;

\caption{$CONSTRUCTION$($V_1, \cdots, V_k$)}
\label{pF2Construction}
\end{scriptsize}
\end{procedure}
\end{spacing}
\vspace*{1pt}


\begin{lemma}
The time complexity of Procedure \ref{pF2Construction} is $O(kn^2)$.
\label{lConstruction}
\end{lemma}
\paragraph*{Proof}
The procedure  has only one loop in which the algorithm from \cite{pacheco2020-cii} is 
invoked $k$ times. Since the number of vertices in each tour is trivially bounded by $n$, 
the time complexity of the algorithm is $O(kn^2)$.

\section{Phase 3: Improvement}
\label{sImprovement}

In Phase 3, the improvement of the feasible solution of Phase 2 is 
carried out in two stages. In stage 1, a sub-tour of some 
tour is repeatedly determined and inserted to some other tour. 
Stage 2 iteratively relocates a vertex from a tour within the same
or another tour, and it may also interchange two vertices from different tours.
We need to following definitions to describe Phase 3 in detail. 

The {\it gain} $g(j, p, m)$ specifies the reduction of the cost 
of tour $T^j= (d, \cdots, i^j_p, i^j_{p+1}, \cdots, i^j_{p+m-1}, \cdots, d)$ \sloppy when 
its subtour $(i^j_p, \cdots, i^j_{p+m-1})$ is removed from that tour. Formally, 
\begingroup
\small
\setlength{\abovedisplayskip}{1pt}
\setlength{\belowdisplayskip}{1pt}
\begin{equation}
g(j,p,m)=w(i^j_{p-1},i^j_{p+m}) - w(i^j_{p-1},i^j_{p}) - \sum_{\iota:=0}^{m-2} w(i^j_{p+\iota}, i^j_{p+\iota+1}) - w(i^j_{p+m-1},i^j_{p+m})
\end{equation}
\endgroup

\begin{spacing}{1.0}
\begin{figure}[H] 
\setlength{\belowcaptionskip}{-15pt}
\begin{center}
 \begin{tabular}[b]{c}
 \includegraphics[width=.22\linewidth]{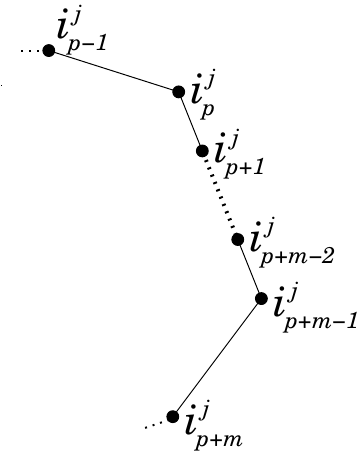} \\
 \small (\textbf{a}) A sub-tour \\ $(i^j_p, \cdots, i^j_{p+m-1})$ \\ 
 \small included in $T^j$
 \end{tabular}
 \begin{tabular}[b]{c}
 \includegraphics[width=.22\linewidth]{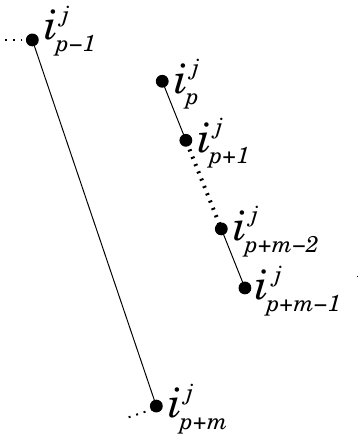} \\
 \small (\textbf{b}) Tour $T^j$ after \\ removing the sub-tour \\
 \small $(i^j_p, \cdots, i^j_{p+m-1})$.
 \end{tabular}
 \caption{Example of removing of a sub-tour}
 \label{fgain}
\end{center}
\end{figure}
\end{spacing}

\begin{figure}[H]    
\setlength{\belowcaptionskip}{-30pt}
\begin{center}
 \begin{tabular}[b]{c}
 \includegraphics[width=.22\linewidth]{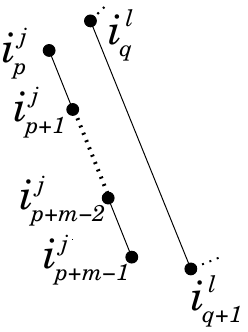} \\ ~ \\
 \scriptsize (\textbf{a}) Sub-tour $(i^l_q, i^l_{q+1})$\\
 \scriptsize and the sub-tour\\
 \scriptsize $(i^j_p, \cdots, i^j_{p+m-1})$ \\ ~
 \end{tabular}
 \begin{tabular}[b]{c}
 \includegraphics[width=.22\linewidth]{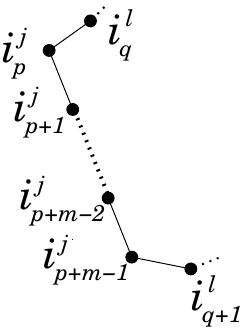} \\
 \scriptsize (\textbf{b}) Sub-tour\\
 \scriptsize $(i^l_q, i^j_p, \cdots, i^j_{p+m-1}, i^l_{q+1})$ \\
 \scriptsize after the insertion of\\
 \scriptsize $(i^j_p, \cdots, i^j_{p+m-1})$\\
 \scriptsize in the directipon $\rho=1$
 \end{tabular}
 \begin{tabular}[b]{c}
 \includegraphics[width=.22\linewidth]{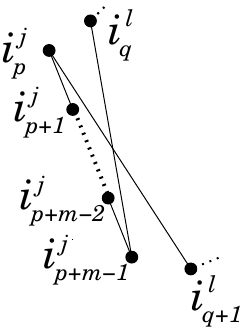} \\
 \scriptsize (\textbf{c}) Sub-tour\\
 \scriptsize $(i^l_q, i^j_{p+m-1}, \cdots, i^j_p, i^l_{q+1})$\\
 \scriptsize after the insertion of\\
 \scriptsize $(i^j_p, \cdots, i^j_{p+m-1})$\\ 
 \scriptsize in the direction $\rho=-1$
 \end{tabular}
 \caption{Example of inserting a sub-tour $(i^j_p, \cdots, i^j_{p+m-1})$ with $m$ consecutive vertices into tour $T^l$}
 \label{fcost}
\end{center}
\end{figure}

Now we incorporate sub-tour $(i^j_p, \cdots, i^j_{p+m-1})$ in tour $T^l$
between two adjacent vertices, say $(i^l_q$ and
$ i^l_{q+1}$ of that tour (see Fig. \ref{fcost}.a).
We apply two alternative ways for the insertion, in the forward and the 
backward fashions, see Fig. \ref{fcost}.b and \ref{fcost}.c (we will use 
auxiliary function $\rho$ with $\rho=1$ for the forward and $\rho=-1$ for 
the backward insertions). Correspondingly, the cost $c( (i^j_p, \cdots, i^j_{p+m-1}), l, q, \rho)$ for the insertion of an sub-tour 
$(i^j_p, \cdots, i^j_{p+m-1})$ in  $T^l$ between the vertices $i^l_q$ and $i^l_{q+1}$ is
\begingroup
\scriptsize
\setlength{\abovedisplayskip}{1pt}
\setlength{\belowdisplayskip}{1pt}
\begin{multline*}
c\left( (i^j_p, \cdots, i^j_{p+m-1}), l, q,\rho\right)=
\left\{ \begin{aligned}
w(i^l_q, i^j_p) + \sum_{\iota:=0}^{m-2} w(i^j_{p+\iota}, i^j_{p+\iota+1}) + w(i^j_{p+m-1}, i^l_{q+1}) - w(i^l_q, i^l_{q+1}) \ \ \mathrm{\mathbf{if}} \ \rho=1 \\
w(i^l_q, i^j_{p+m-1}) + \sum_{\iota:=0}^{m-2} w(i^j_{p+\iota}, i^j_{p+\iota+1}) + w(i^j_p, i^l_{q+1}) - w(i^l_q, i^l_{q+1}) \ \  \mathrm{\mathbf{if}} \ \rho=-1
\end{aligned}  \right .
\end{multline*}
\endgroup

\begin{spacing}{1.0}
\end{spacing}
\unskip

We perform either the forward or the backward insertion, one with the minimum
cost, determined as follows.
\begingroup
\small
\setlength{\abovedisplayskip}{1pt}
\setlength{\belowdisplayskip}{1pt}
$$
\rho(j, p, m, l, q)= 
\left\{ \begin{aligned}
1 \ \ \ &\mathrm{\mathbf{if}} \  w(i^l_q, i^j_p)+w(i^j_{p+m-1}, i^l_{q+1}) 
& \leq  w(i^l_q, i^j_{p+m-1})+w(i^j_p, i^l_{q+1})\\
-1 \ \ \ &\mathrm{\mathbf{otherwise}} \ 
\end{aligned}  \right .
$$
\endgroup

In the formal descriptions, we use an auxiliary global  variable $g$ that keeps track 
of the gain  of relocation/swapping, accomplished in each of the procedures. The 
parameter $g$ is used in our halting conditions.

\subsection{Procedure RELOCATE\_SUBTOURS}
\label{sRelocateSubtours}

Now we describe an iterative procedure that relocates a sub-tour of a tour
to another tour until the total cost cannot be reduced. An {\em admissible}
sub-tour of a tour is one that contains no more than $\eta_{\max}$ 
vertices, the minimum between
the maximum number of vertices that can be removed from a tour $T^j$, 
$( \ \max_{j=1,\cdots,k} \{ m_j-m_{\min} \} \ )$ and
the maximum number of vertices that can be received by a tour $T^l$, 
$( \ \max_{l=1,\cdots,k} \{ m_{\max}-m_{l} \} \ )$, where $m_j=|T_j|$:
\begingroup
\small
\setlength{\abovedisplayskip}{1pt}
\setlength{\belowdisplayskip}{1pt}
$$\eta_{\max}:=\min \left\{ \ \max_{j=1,\cdots,k} \{ m_j-m_{\min} \} \ , \ \max_{l=1,\cdots,k} \{ m_{\max}-m_l \} \ \right\}$$
\endgroup
We verify every admissible sub-tour of every tour $T_j$. For each admissible 
sub-tour $(i^j_p, i^j_{p+1}, \cdots, i^j_{p+m-1})$ of tour $T^j$,
$2 \leq m \leq \eta_{\max}$, $m_j-m \geq m_{\min}$ and $1 \leq p \leq m_j-m+1$,
the gain 
\begingroup
\small
\setlength{\abovedisplayskip}{1pt}
\setlength{\belowdisplayskip}{1pt}
$$g(j,p,m)+c\left((i^j_p, i^j_{p+1}, \cdots, i^j_{p+m-1}), l, q, \rho(j, p, m, l, q) \right)$$
\endgroup
of  its relocation to tour $T^l$ in 
between the vertices $i^l_q$ and $i^l_{q+1}$,
$m_l+m \leq m_{\max} $ and $0 \leq q \leq m_l$ is calculated. 

In case none of the gains are negative, the procedure halts. Otherwise, the
relocation is accomplished for any sub-tour with the minimum gain. Then a local 
search algorithm 2-OPT \cite{croes1958} is carried out for the two modified
tours.  

\begin{spacing}{1.0}
\begin{procedure}[ht]   
\begin{scriptsize}
\SetKwInOut{Input}{input}
\DontPrintSemicolon	

$\eta_{\max}:=\min \left\{ \ \max_{j=1,\cdots,k} \{ m_j-m_{\min} \} \ , \ \max_{l=1,\cdots,k} \{ m_{\max}-m_l \} \ \right\}$\;
\While{$g<0$}{
 $g:=\infty$\;
 \For {$m:=2$ \KwTo $\eta_{\max}$}{
  \For {$j:=1$ \KwTo $k$}{
   \If {$m_j-m \geq m_{\min}$}{
    \For {$p:=1$ \KwTo $m_j-m+1$}{
     \For {$l:=1$ \KwTo $k$}{
      \If {$ m_l+m \leq m_{\max} $}{
       \For {$q:=0$ \KwTo $m_l$}{
        \If{$j \neq l \wedge g(j,p,m)+$ 
          $c((i^j_p, \cdots, i^j_{p+m-1}), l, q, $
          $ \rho(j, p, m, l, q) )$
         $< g$}{
         $g:=g(j,p,m)+$
          $c((i^j_p, \cdots, i^j_{p+m-1}), l, q, $
          $ \rho(j, p, m, l, q) )$ \;
         $\mu:=m$, $\iota:=j$, $\ \alpha:=p$ \;
         $\varrho:=\rho(j, p, m, l, q)$, $\ \lambda:=l$, $\beta:=q$\;
        }
       }
      }
     }
    }
   }
  }
 } 
 
 \If{$g<0$}{
  $T^{*}:=(i^\iota_\alpha, \cdots, i^\iota_{\alpha+\mu-1})$\;
  $T^\iota:=$ Remove subtour $T^{*}$ from tour $T^\iota$\;
  $T^\lambda:=$ Insert subtour $T^{*}$ into tour $T^\lambda$ in direction $\varrho$ between vertices $i^\lambda_\beta$ and $i^\lambda_{\beta+1}$ \;
  $T^\iota:=$2-OPT$(T^\iota)$\;
  $T^\lambda:=$2-OPT$(T^\lambda)$\;
 }
}

\Return $T$ \;

\caption{$RELOCATE_-SUBTOURS$($T$)}
\label{pRelocateSubtours}
\end{scriptsize}
\end{procedure}
\end{spacing}
\vspace*{1pt}

\begin{lemma}
The time complexity of Procedure \ref{pRelocateSubtours} is $O(k^2 n^4)$.
\label{lRelocateSubtours}
\end{lemma}
\paragraph*{Proof}
In lines 2-20, the loop is repeated less than $n$  times.
In lines 4-14, the loop is repeated less than $m_{\max} n$ times.
Since $m_{\max}$ is bounded by $n$, the cost for this loop is $O(n^2)$.
In lines 5-14, the loop is repeated less than $m_{\max} k n$ times, 
therefore the  cost is $O(k n^2)$.
In lines 7-14, the loop is repeated less than $m^2_{\max} k n$ times, 
therefore the cost is $O(k n^3)$.
In lines 8-14, the loop is repeated less than $m^2_{\max} k^2 n$ times, 
therefore the cost is $O(k^2 n^3)$.
In lines 10-14, the loop is repeated less than $m^3_{\max} k^2 n$ times, 
therefore the cost is $O(k^2 n^4)$.
In lines 19 and 20, the 2-OPT algorithm is applied, which has a time complexity 
of $O(n^2)$. 
This algorithm is invoked no more than $n$ times yielding cost $O(n^3)$
and the lemma follows.

Based on our experimental study,  the average number of repetitions in 
the above loops (and also in the loop in Procedure MAIN of Section \ref{sMain})
turned out to be about $\frac{1}{8}n$. Hence, $n$ is a bit rough estimation
for the practical running times for these loops. 


\subsection{Procedure RELOCATE\_A\_VERTEX}
\label{sRelocateAVertex}

This procedure accomplishes a vertex relocation either within the  tour to which 
it belongs or to another tour. A relocation resulting in a new feasible solution 
with the minimum (negative) gain is chosen. For that, for each tour $T^j$ 
and each vertex $i^j_p$, $m_j > m_{\min}$ and $1 \leq p \leq m_j$, the gain  
\begingroup
\small
\setlength{\abovedisplayskip}{1pt}
\setlength{\belowdisplayskip}{1pt}
$$g(j,p,1)+c\left((i^j_p), l, q, 1\right)$$ 
\endgroup
of its relocation to
tour $T^l$ in between vertices $i^l_q$ and $i^l_{q+1}$,
$ m_l < m_{\max} $,  $0 \leq q \leq m_l$ is calculated. 
A local search algorithm 2-OPT for the two modified tours is similarly carried out.\\

\begin{spacing}{1.0}
\begin{procedure}[ht] 
\begin{scriptsize}
\SetKwInOut{Input}{input}
\DontPrintSemicolon	

$g:=\infty$\;
 \For {$j:=1$ \KwTo $k$}{
  \If {$m_j > m_{\min}$}{
   \For {$p:=1$ \KwTo $m_j$}{
    \For {$l:=1$ \KwTo $k$}{
     \If {$ m_l < m_{\max} $}{
      \For {$q:=0$ \KwTo $m_l$}{
       \If{$i^j_p \neq i^l_q \wedge 
           i^j_p \neq i^l_{q+1} \wedge
       g(j,p,1)+c\left((i^j_p), l, q, 1\right)< g$}{
        $g:=g(j,p,1)+c\left((i^j_p), l, q, 1\right)$\;
        $\iota:=j$, $\alpha:=p$\;
        $\lambda:=l$, $\beta:=q$\;
       }
      }
     }
    }
   }
  }
 }
 \eIf{$g<0$}{
  $i:=i^\iota_\alpha$\;
  $T^\iota:=$ Remove vertex $i$ from tour $T^\iota$\;
  $T^\lambda:=$ Insert vertex $i$ into tour $T^\lambda$ between vertices $i^\lambda_\beta$ and $i^\lambda_{\beta+1}$ \;
  $T^\iota:=$2-OPT$(T^\iota)$\;
  $T^\lambda:=$2-OPT$(T^\lambda)$\;
 }{
  $g:=0$\;
 }

\Return $T$ \;

\caption{$RELOCATE_-A_-VERTEX$($T$)}
\label{pRelocateAVertex}
\end{scriptsize}
\end{procedure}
\end{spacing}

\begin{lemma}
The time complexity of Procedure \ref{pRelocateAVertex} is $O( k^2 n^2 )$.
\label{lRelocateAVertex}
\end{lemma}
\paragraph*{Proof}
In lines 2-11, the loop is repeated $k$ times.
In lines 4-11, the loop is repeated $m_{\max} k$ times,
Since $m_{\max}$ is bounded by $n$, the cost for this loop is $O(k n)$.
In lines 5-11, the loop is repeated $m_{\max} k^2$ times,
therefore the  cost is $O(k^2 n)$.
In lines 7-11, the loop is repeated $(m_{\max}-1) m_{\max} k^2= m^2_{\max} k^2 - m_{\max} k^2$ times,
therefore the cost is $O(k^2 n^2)$.
In lines 16 and 17, the 2-OPT algorithm is applied, which has a time complexity 
of $O(n^2)$. Hence this algorithm yields a cost of $O(n^2)$
and the lemma follows.

\subsection{Procedure SWAP\_VERTICES}
\label{sSwapVertices}

This procedure swaps two vertices of different tours. So let $i^j_p$ be a vertex in  tour $T^j$ and let $i^l_q$ be a vertex in tour $T^l$ in current solution $T$, 
$j \neq l$. For every this pair of vertices,  the gain 
\begingroup
\small
\setlength{\abovedisplayskip}{1pt}
\setlength{\belowdisplayskip}{1pt}
$$g(j,p,1)+c\left((i^j_p), l, q, 1\right)+ g(l,q,1)+c\left((i^l_q), j, p, 1\right)$$ 
\endgroup
obtained by swapping position within solution  $T$ is calculated.
A swap with the minimum (negative) gain performed out and again, the 
2-OPT algorithm is carried out for the two modified tours. \\

\begin{spacing}{1.0}
\begin{procedure}[ht]  
\begin{scriptsize}
\SetKwInOut{Input}{input}
\DontPrintSemicolon	

$g^{\prime}:=\infty$\;

 \For {$j:=1$ \KwTo $k-1$}{
  \For {$p:=1$ \KwTo $m_j$}{
   \For {$l:=j+1$ \KwTo $k$}{
    \For {$q:=1$ \KwTo $m_l$}{

     \If{$g(j,p,1)+c\left((i^j_p), l, q, 1\right)+ g(l,q,1)+c\left((i^l_q), j, p, 1\right)<g^{\prime}$}{
        $g^{\prime}:=g(j,p,1)+c\left((i^j_p), l, q, 1\right)+ g(l,q,1)+c\left((i^l_q), j, p, 1\right)$\;
        $\iota:=j$, $\alpha:=p$\;
        $\lambda:=l$, $\beta:=q$\;
     }
    }
   }
  }
 }

 \If{$g^{\prime}<0$}{
  Swap the vertices $i^\iota_\alpha$ and $i^\lambda_\beta$ from $T^\iota$ and $T^\lambda$ respectively\;
  $T^\iota:=$2-OPT$(T^\iota)$\;
  $T^\lambda:=$2-OPT$(T^\lambda)$\;
  $g:=g+g^{\prime}$\;
 }
 \Return $T$ 

\caption{$SWAP_-VERTICES$($T$)}
\label{pSwapVertices}
\end{scriptsize}
\end{procedure}
\end{spacing}

\begin{lemma}
The time complexity of Procedure \ref{pSwapVertices} is $O( k^2 n^2 )$.
\label{lSwapVertices}
\end{lemma}
\paragraph*{Proof}
In lines 2-9, the loop is repeated $k-1$ times.
In lines 3-9, the loop is repeated $m_{\max} (k-1)$.
Since $m_{\max}$ is bounded by $n$, the computational cost for this loop is $O(k n)$.
In lines 4-9, the loop is repeated $m_{\max} (k-1)^2$ times,
therefore the computational cost is $O(k^2 n)$.
In lines 5-9, the loop is repeated $m_{\max}^2 (k-1)^2$ times,
therefore the computational cost is $O(k^2 n^2)$.
In lines 12 and 13, the 2-OPT algorithm is applied, which has a time complexity 
of $O(n^2)$ and the lemma follows.

\section{Procedure MAIN}
\label{sMain}

Below we give a formal description of the overall algorithm incorporating the earlier described procedures. In summary, for the partition of the set of vertices in $k$ subsets, 
constructed in procedure PARTITION(), procedure CONSTRUCTION($V_1$, $\cdots$, $V_k$)
creates an initial solution $T_0$, which is transformed to solution $T_1$ in 
procedure RELOCATE\_SUBTOURS. Further improvements are done in 
procedures RELOCATE\_A\_VERTEX and SWAP\_VERTICES, which are alternately invoked. 

\medskip

\begin{spacing}{1.0}
\begin{algorithm}[ht]
\begin{scriptsize}

\SetKwInOut{Input}{input}
\DontPrintSemicolon

$g:=-1$\;

$V_1, \cdots, V_k:=PARTITION()$ \tcp*[r]{Phase 1}

$T_0:=CONSTRUCTION(V_1, \cdots, V_k)$ \tcp*[r]{Phase 2}

$T_1:=RELOCATE_-SUBTOURS(T_0)$ \tcp*[r]{Phase 3}

$h:=1$\;
$g:=-1$\;
\While{$g<0$}{
 $T_{h+1}:=RELOCATE_-A_-VERTEX(T_h)$\;
 $h:=h+1$\;
 $T_{h+1}:=SWAP_-VERTICES(T_h)$\;
 $h:=h+1$\;
}

\Return $T_h$ \;

\caption{$MAIN$($V, k, m_{\min}, m_{\max}$)}
\label{pMAIN}
\end{scriptsize}
\end{algorithm}
\end{spacing}
\vspace*{1pt}

\begin{lemma}
The worst-case time complexity of the algorithm is $O(k^2 n^4)$.
\end{lemma}
\paragraph*{Proof} ~ 
The outer loop and the loop in lines 7-11 in procedure \ref{pMAIN}  is repeated
less than $n$ times. 
By Lemmas \ref{lPartition}, \ref{lConstruction}, \ref{lRelocateSubtours}, \ref{lRelocateAVertex} and \ref{lSwapVertices} the
time complexities of procedures  {\em PARTITION},  {\em CONSTRUCTION}, {\em RELOCATE\_SUBTOURS}, {\em RELOCATE\_A\_VERTEX} and {\em SWAP\_VERTICES} are 
$O(k n^3)$,  $O(k n^2)$, $O(k^2 n^4)$, $O( k^2 n^2 )$ and $O( k^2 n^2 )$ respectively. 
Lines 7-11 yield time $O( k^2 n^3 )$ and the overall time complexity is 
$O(k^2 n^4)$

\section{The implementation and results of our experiments}
\label{sResults}

PCI-algorithm was coded in C++ and compiled in g++ on a server with processor 
2x Intel Xeon E5-2650 0 @ 2.8 GHz, 32 GB in RAM and GNU/Linux Debian 12 (bookworm) operating system (we  used only one CPU in our experiments). We did not keep the cost matrix in computer memory, we were rather calculating the costs using the coordinates of the points. This did not increase the computation time too much but saved considerably computer memory. For instance, 
for the pr439\_5 instance, only 785 KiB of memory were used, pr1002\_5 consumed 1 MiB, and the largest instance, fl1577\_42, used 1.2 MiB of memory.

The detailed experimental data including the program code, our 168 new instances, the obtained solutions and analysis of our results can be accessed at \cite{pacheco2023-benchmarks}. 


Next, we describe how we created our new problem instances available at
\cite{pacheco2023-benchmarks}). They were derived from 56 benchmark TSP instances 
from TSPLIB \cite{TSPLIB-Benchmarks}.
For each of these instances, the vertex with index 1  was set as the depot in the
corresponding BMTSP instances. We created three BMTSP problem instances for each 
of the TSP instances by setting three kits of  parameters $k$,  $m_{\min}$   
and $m_{\max}$, as follows. The three chosen values for $m_{\max}$ were $30$, 
$40$, and $50$, which were determined based on a private communication 
with several logistic companies (as an example, 
computer technicians from a convenience store chain visit up to $30$ stores in  
nearby towns  to attend simple repairs in the point-of-sale system;  
snack distributors visit up to $30$ clients in different municipalities; 
soft drink distributors visit between $40$ and $50$ stores in a city;
snack distributors deliver and arrange their products in $16$ to $20$ small 
shops of various sizes). For each so determined $m_{\max}$, the 
corresponding  $m_{\min}$ was set to $0.6$ times  $m_{\max}$. This selection 
of  $m_{\min}$ was also motivated by practice (in practice, a salesman 
is assigned such a reasonable number of clients to justify his/her wage). 
With the so determined bounds, we set the number of salesmen to $\left \lceil 1.3n/m_{\max} \right \rceil$ (such a 
distribution of $n$ cities in  $k$ tours  ensures that the lower and upper bounds 
are met as per the specified restrictions). 

\begin{table}[H]  
\caption{The summary of results for the 22 benchmarks instances from \cite{junjie2006, necula2015}}
\label{tComparison}
\centering
\resizebox{0.8\textwidth}{!}{%
\begin{tabular}{lrlrrrr}   
\hline
\multicolumn{1}{c}{\textbf{Instance}} & \multicolumn{1}{c}{$\mathbf{C(BKS)}$} & \multicolumn{1}{c}{\textbf{Author}} & \multicolumn{1}{c}{\textbf{Time$_{BKS}$}} & \multicolumn{1}{c}{$\mathbf{C(PCI)}$} & \multicolumn{1}{c}{\textbf{Time$_{PCI}$}} & \multicolumn{1}{c}{\textbf{GAP}} \\
\hline
pr76\_5 & 153,389.90 & \cite{lo2018} & 354 s & \textbf{151,568.87} & 3.83 s & \textbf{-1.19\%} \\ 
pr152\_5 & 114,752.00 & \cite{alfurhud2020} & NA & \textbf{113,598.83} & 22.40 s & \textbf{-1.00\%} \\
pr226\_5 & 147,586.00 & \cite{alfurhud2020} & NA & \textbf{143,217.88} & 320.20 s & \textbf{-2.96\%} \\ 
pr299\_5 & 72,664.00 & \cite{alfurhud2020} & NA & \textbf{69,574.27} & 827.31 s & \textbf{-4.25\%} \\ 
pr439\_5 & 138,527.00 & \cite{alfurhud2020} & NA & \textbf{134,471.63} & 1,012.06 s & \textbf{-2.93\%} \\ 
pr1002\_5 & 329,128.00 & \cite{pacheco2021-ioca}& 22 s & \textbf{305,417.67} & 29,200.62 s & \textbf{-7.20\%} \\ 
\hline
eil51\_2 & 442.32$^{\mathrm{a}}$ & \cite{necula2015} & NA & 442.71 & 0.07 s & 0.09\% \\
eil51\_3 & 464.11$^{\mathrm{a}}$ & \cite{necula2015} & NA & 464.11 & 1.98 s & 0.00\% \\
eil51\_5 & 529.70 & \cite{necula2015}& NA & 529.70 & 3.57 s & 0.00\% \\
eil51\_7 & 605.21 & \cite{necula2015} & NA & \textbf{601.99} & 3.43 s & \textbf{-0.53\%} \\
berlin52\_2 & 7,753.89$^{\mathrm{a}}$ & \cite{necula2015} & NA & 7,753.89 & 0.86 s & 0.00\% \\
berlin52\_3 & 8,106.85$^{\mathrm{a}}$ & \cite{necula2015} & NA & 8,106.85 & 5.94 s & 0.00\% \\
berlin52\_5 & 9,126.33 & \cite{necula2015} & NA & \textbf{9,118.29} & 9.91 s & \textbf{-0.09\%} \\
berlin52\_7 & 9,870.02 & \cite{necula2015} & NA & \textbf{9,808.26} & 10.20 s & \textbf{-0.63\%} \\
eil76\_2 & 558.59$^{\mathrm{a}}$ & \cite{necula2015} & NA & 570.35 & 0.05 s & 2.11\% \\
eil76\_3 & 579.30$^{\mathrm{a}}$ & \cite{necula2015} & NA & 581.20 & 12.59 s & 0.33\% \\
eil76\_5 & 680.67 & \cite{necula2015} & NA & \textbf{676.81} & 2.93 s & \textbf{-0.57\%} \\
eil76\_7 & 759.90 & \cite{necula2015} & NA & \textbf{749.61} & 16.14 s & \textbf{-1.35\%} \\ 
rat99\_2 & 1,350.73 & \cite{necula2015} & NA & \textbf{1,340.39} & 2.65 s & \textbf{-0.77\%} \\ 
rat99\_3 & 1,519.49 & \cite{necula2015} & NA & \textbf{1,518.72} & 15.73 s & \textbf{-0.05\%} \\ 
rat99\_5 & 1,855.83 & \cite{necula2015} & NA & \textbf{1,812.40} & 2.35 s & \textbf{-2.34\%} \\ 
rat99\_7 & 2,291.82 & \cite{necula2015} & 14 h & \textbf{2,204.27} & 2.14 s & \textbf{-3.82\%} \\ 
\hline
\multicolumn{5}{l}{$^{\mathrm{a}}$Reported as optimal.}
\end{tabular}
} 
\end{table}
\unskip

Table \ref{tComparison} presents the results obtained by our algorithm for the 
22 benchmark instances. The ``Instance'' column lists the  
abbreviated names of the instances, the column ``$C(BKS)$'' displays the cost 
of the BKS, and the column ``Time$_{BKS}$'' gives the time at which 
the corresponding algorithm achieved the BKS (whenever this time is available). 
The ``$C(PCI)$'' column shows the cost of the best obtained solution, 
and the "Time$_{PCI}$" column shows the time that our algorithm required 
to create this solution. The column ``GAP''  indicates a relative percentage error 
between $C(BKS)$ and $C(PCI)$  defined as 
\begingroup
\setlength{\abovedisplayskip}{2pt}
\setlength{\belowdisplayskip}{2pt}
\begin{equation}
GAP(PCI,BKS)=\frac{(C(PCI) - C(BKS))}{C(BKS)} \times 100\%
\label{eGap}
\end{equation}
\endgroup
(note that PCI-algorithm improves the earlier BKS for the entries marked in bold, 
i.e., the negative entries in Table \ref{tComparison}).

As we can see from Table \ref{tResults} below, PCI-algorithm improved
the earlier known best costs for the 15 from the 22 benchmark 
instances (12 of them being the largest ones). It is remarkable that
for the problem instance ``pr76\_5'' the solution obtained by our algorithm
was further improved by CPLEX. For the 4 benchmark instances, 
the solutions obtained by our algorithm have the same cost as the earlier 
BKS. For the remaining 3 small instances, the ones between 51 and 76 cities 
and with 2 or 3 salesmen, where optimal solutions were earlier obtained by CPLEX, 
PCI-algorithm gave the approximation gap of 0.09\%, 2.11\% and 0.33\%.

\begin{table}[H]  
\caption{Quality of our solutions compared to their BKS for the 22 benchmark instances}
\label{tResults}
\centering
\resizebox{0.6\textwidth}{!}{%
\begin{tabular}{|r|l|}
\hline
\textbf{Number} & \textbf{Description}\\
\hline
14 & Solutions with lower cost than their BKS\\ 
1 & Solution obtained by CPLEX with lower cost than their BKS\\ 
3 & Solutions equal to optimal solution\\ 
1 & Solution equal than their BKS\\ 
3 & Solutions with an error between 0\% and 3\%\\ 
\hline
\end{tabular} 
}
\end{table}

\begin{figure}[H]  
\begin{center}
 \begin{tabular}[b]{c}
 \includegraphics[width=.5\linewidth]{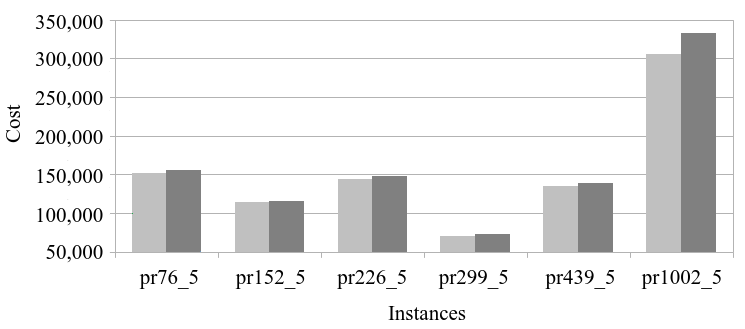}\\
 \footnotesize (a) Instances from \cite{junjie2006}
 \vspace{0.2 cm}
 \end{tabular}

 \begin{tabular}[b]{c}
 \includegraphics[width=0.35\linewidth]{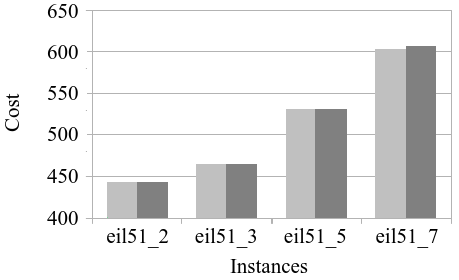}\\ 
 \footnotesize (b) Instances derived from eil51 from \\ 
 \footnotesize \cite{necula2015}
  \vspace{0.2 cm}
 \end{tabular}
 \begin{tabular}[b]{c}
 \includegraphics[width=.35\linewidth]{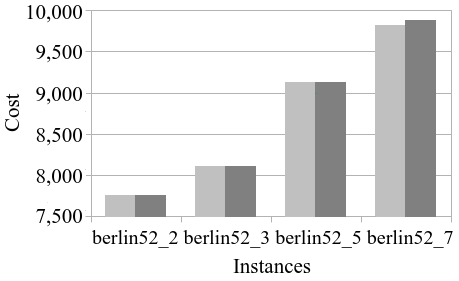}\\
 \footnotesize (c) Instances derived from berlin52 \\
 \footnotesize \cite{necula2015}
 \vspace{0.2 cm}
 \end{tabular}

 \begin{tabular}[b]{c}
 \includegraphics[width=0.35\linewidth]{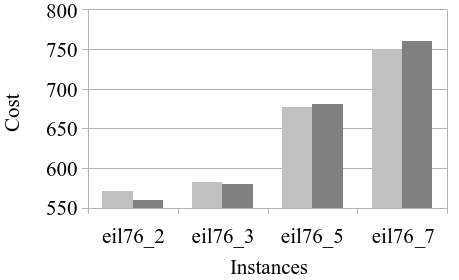}\\
 \footnotesize (d) Instances derived from eil76 from \\ 
 \footnotesize \cite{necula2015}
 \vspace{0.2 cm}
 \end{tabular}
 \begin{tabular}[b]{c}
 \includegraphics[width=.35\linewidth]{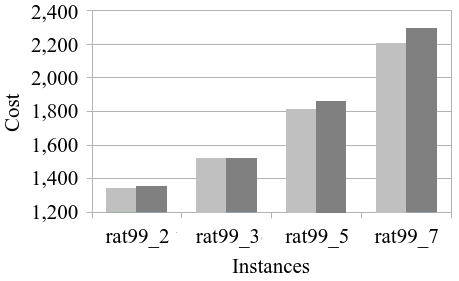}\\
 \footnotesize (e) Instances derived from rat99 from \\ 
 \footnotesize \cite{necula2015}
 \vspace{0.2 cm}
 \end{tabular}
 \begin{tabular}[b]{c}
 \includegraphics[width=.45\linewidth]{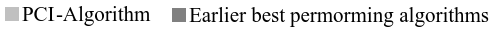}
 \end{tabular}
 \caption{Costs for the 22 benchmarks instances from \cite{junjie2006,necula2015}}
 \label{fCost22}
\end{center}
\end{figure}

Charts from Figure \ref{fCost22} compare the costs obtained by PCI-algorithm 
vs earlier known best costs  for the 6 instances from \cite{junjie2006} 
and for 16 instances with the earlier obtained results by CPLEX, as 
reported in  \cite{necula2015}.  

\unskip

\begin{figure}[H] 
\setlength{\belowcaptionskip}{-20pt}
\begin{center}
 \begin{tabular}[b]{c}
 \includegraphics[width=0.45\linewidth]{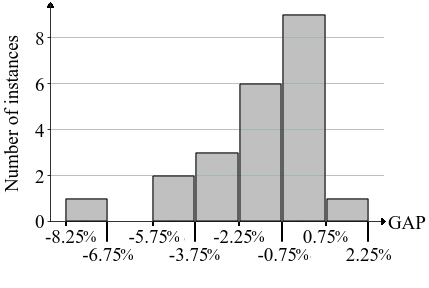}\\
 \footnotesize (\textbf{a}) Histogram
 \end{tabular}
 \begin{tabular}[b]{c}
 \includegraphics[width=.46\linewidth]{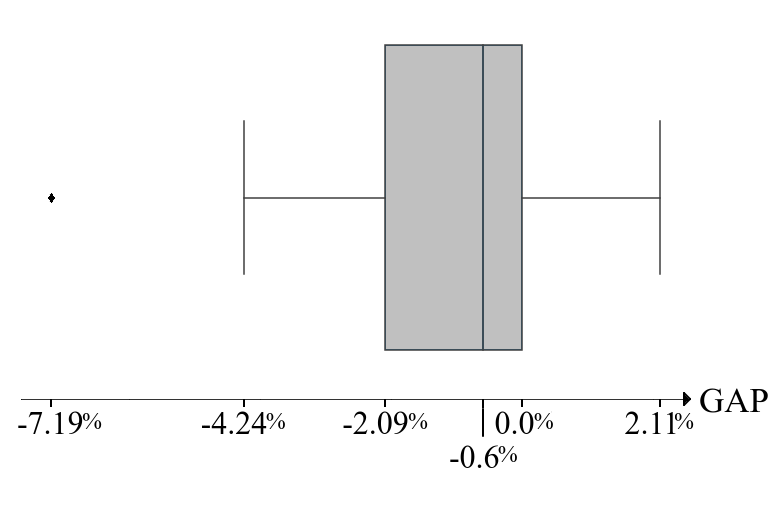}\\
 \footnotesize (\textbf{b}) Box plot
 \end{tabular}
 \caption{Quality of our solutions compared to their BKS for the 22 benchmark instances}
  \label{fstatistics}
\end{center}
\end{figure}

In the histogram from Figure \ref{fstatistics}, the x-axis represents different
$GAP(PCI,BKS)$ ranges (see formula \ref{eGap}) divided into 7 intervals. 
The y-axis indicates the number of instances that fall within each of these intervals. 
For instance, the interval with limits of -0.75\% and 0.75\% has the highest count 
of 9 instances, followed by intervals with limits between -2.25\% and -0.75\%, -3.75\% 
and -2.25\%, and -5.75\% and -3.75\%.  
Figure \ref{fstatistics}.b displays a box plot illustrating the $GAP(PCI,BKS)$
distribution (see Formula \eqref{eGap}) for the 22 benchmarks instances.
The central box in the plot represents the interquartile range, between the first quartile and the third quartile, covering 50\% of the instances. These values fall between -2.09\% and 0.00\% of the $GAP(PCI,BKS)$ (the median, represented by the line within the box, is  -0.6\%).
The outlier to the left of  
the lower whisker at -7.20\% corresponds to instance pr1002\_5.

\begin{figure}[H]  
\setlength{\abovecaptionskip}{-2pt}
\setlength{\belowcaptionskip}{-10pt}
\begin{center}
 \includegraphics[width=0.8\linewidth]{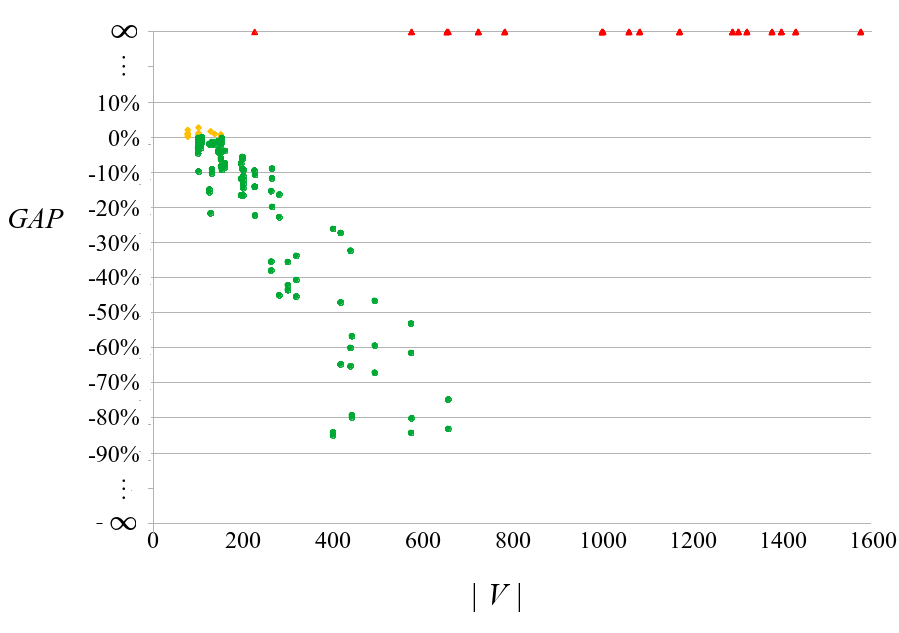}
 \caption{$GAP(PCI,CPLEX)$ between CPLEX and PCI-algorithm for the 168 new instances}
 \label{fErrorCPLEX}
\end{center}
\end{figure}
\unskip

In figure \ref{fErrorCPLEX} we can observe the dependence of $GAP(PCI,CPLEX)$ 
(see equation \ref{eGap}) on the instance sizes. There are 12 instances 
with size less than 100, where the $GAP(PCI,CPLEX) \in[ 0\%,2.67\%]$, and 
5 instances with size in the range $[100,150]$ 
with $GAP(PCI,CPLEX)$ 0\% and 1.61\% (yellow dots). For the 
remaining instances, $GAP(PCI,CPLEX) < 0$, i.e.,  PCI-algorithm gave better solutions 
(green dots). For 49  instances with more than 574 vertices
CPLEX could not provide a feasible solution within the
imposed by us time limit of two hours or/and because it ran out of 
available memory (red dots). 

\section{Conclusions}
\label{sConclusions}

The proposed PCI-algorithm for the Euclidean BMTSP obtained BKS for 19 of the 
22 benchmark instances,  including the twelve largest instances. The 
remaining 3 small benchmark instances were solved optimally  by CPLEX.
168 new benchmark instances were created, that are now publicly available. 
For the  17 smallest of them, CPLEX obtained better solutions (with  the maximum 
$GAP(PCI,BKS)$ of 2.67\% and average $GAP(PCI,BKS)$ of 0.86\%), 
PCI-algorithm was better for 102 instances (with the maximum
$GAP(PCI,BKS)$ of -85\% and average $GAP(PCI,BKS)$ of -21.42\%), whereas
CPLEX failed to deliver feasible solutions for the remaining 49 instances
within the imposed time limit of two hours.

The random choice of the initial vertex of each subset at Phase-1 yielded 
diverse initial solutions at different runs and better overall costs. 
The application of sub-tour relocation in our hill climbing 
procedure \ref{pRelocateSubtours} effectively reduced costs of initial solutions. 
For future work, it would be beneficial to further speed up procedure \ref{pRelocateSubtours} enhancing phase 3 with faster local search procedures.  
Another natural direction is the extension of PCI-algorithm for other variations 
of MTSP including capacitated setting and the one with time windows.





\bibliographystyle{apalike} 
\bibliography{mtsp.bib}

\end{document}